\documentclass[12pt,letterpaper]{article}
\usepackage[a4paper, total={7in, 10in}]{geometry}

\usepackage{graphicx}
\usepackage{svg}
\usepackage{helvet}
\usepackage{authblk}
\usepackage{hyperref}
\usepackage{amsmath} 
\usepackage{amssymb} 
\usepackage{orcidlink} 
\usepackage{booktabs}
\usepackage{multirow}
\usepackage{makecell}
\usepackage{placeins}

\usepackage[super,comma,sort&compress]  
   {natbib}
\usepackage{url}

\makeatletter
\renewcommand{\maketitle}{\bgroup\setlength{\parindent}{0pt}
\begin{flushleft}
  \textbf{\@title}
  
  \@author
\end{flushleft}\egroup}
\makeatother

\title{Subcellularly Resolved Single-Cell Embedding Learning with Transcriptomic data, Protein Structure and Localization Information}
\date{}

\author[1]{Zhen Zhou}
\author[2]{Jiachen Li}
\author[1]{Yuan Liu}
\author[1]{Xiaoyong Pan}
\author[1,*]{Hong-Bin Shen}

\affil[1]{Institute of Image Processing and Pattern Recognition, Shanghai Jiao Tong University, and Key Laboratory of System Control and Information Processing, Ministry of Education of China, Shanghai, 200240, China}
\affil[2]{Institute of Process Engineering, Chinese Academy of Sciences, Beijing, 100190‌, China}

\affil[*]{Correspondence: hbshen@sjtu.edu.cn}

\begin{document}

\maketitle

\section*{Abstract}

Existing cell embedding methods predominantly rely on transcriptomic or proteomic measurements and represent each cell as a holistic entity, thereby overlooking the subcellular localization of individual molecules. Moreover, they rarely incorporate protein structural information, despite its fundamental role in determining molecular interactions and functions. In this work, we propose a multimodal framework for learning subcellularly resolved cell embeddings by jointly leveraging RNA expression profiles, protein sequence representations, and protein structural information. Specifically, we employ a cross-attention architecture to integrate transcriptomic, sequence, and structural modalities and model their interactions within distinct subcellular compartments. The resulting embeddings represent each cell through its fine-grained subcellular organization, capturing both molecular expression patterns and the functional properties of the associated proteins. By learning cell representations at subcellular resolution, our framework preserves spatially organized biological information while integrating complementary signals across multiple molecular levels. To the best of our knowledge, this is the first framework
that produces subcellularly resolved cell embeddings by jointly incorporating transcriptomic information, protein sequence representations, and protein structural knowledge within a unified cross-modal learning paradigm.
\section*{Keywords}

Cell Embedding; Subcellular Representation Learning; Single-Cell Transcriptomics; Protein Structure; Protein Sequence; Multimodal Learning; Cross-Attention; Representation Learning
\section*{Introduction}

\section{Introduction}

Recent advances in single-cell RNA sequencing (scRNA-seq) technologies have enabled the systematic profiling of cellular states at unprecedented resolution, giving rise to large-scale cell atlases and foundation models for single-cell analysis. Inspired by the success of large language models, a series of single-cell foundation models have been proposed, including Geneformer \cite{theodoris2023geneformer}, scGPT \cite{cui2024scgpt}, and scFoundation \cite{hao2024large}. These models leverage transformer-based architectures to learn generalizable cellular representations from millions of transcriptomic profiles and have demonstrated strong performance across diverse downstream tasks such as cell type annotation, trajectory inference, and perturbation prediction.

Despite these advances, most existing cell embedding methods remain fundamentally transcriptome-centric. They represent each cell as a collection of gene expression values, implicitly assuming that mRNA abundance is sufficient to capture cellular identity. However, gene expression represents only an intermediate layer of the central dogma and does not directly reflect the functional execution of biological processes. As a result, current cell embeddings often fail to capture key molecular mechanisms underlying cellular behavior.

Proteins are the primary functional molecules in biological systems, directly executing cellular processes through enzymatic activity, molecular interactions, and signaling regulation \cite{alberts2017molecular}. Importantly, transcript abundance alone provides an incomplete description of cellular function, because protein activity is determined not only by expression level but also by three-dimensional structure. Proteins with similar expression patterns may exhibit distinct biological functions due to differences in folding topology, binding interfaces, and structural domains. Recent breakthroughs in protein structure prediction, such as AlphaFold \cite{jumper2021alphafold} and RoseTTAFold \cite{krishna2024generalized}, have revealed that protein structures encode rich functional and evolutionary information that cannot be captured by transcriptomic data alone. Therefore, a biologically meaningful cellular representation should integrate both RNA expression and protein-level information, particularly structural properties that directly determine molecular function. However, existing single-cell foundation models rarely incorporate protein structure, limiting their ability to model cellular function in a mechanistically grounded manner.

In addition to molecular composition, cellular function is also strongly influenced by spatial organization. Eukaryotic cells are highly compartmentalized systems in which biomolecules are dynamically localized across distinct subcellular regions, including the nucleus, cytoplasm, mitochondria, and endoplasmic reticulum. Subcellular localization plays a critical role in regulating gene expression, RNA transport, protein trafficking, and signal transduction \cite{kobayashi2022self}. Moreover, aberrant localization of RNAs and proteins is closely associated with various diseases, highlighting the importance of spatial context in defining cellular states. Nevertheless, most existing cell embedding approaches treat cells as homogeneous entities and neglect intracellular spatial heterogeneity.

Recent advances in imaging technologies, spatial transcriptomics, and large-scale localization datasets have made it increasingly feasible to incorporate subcellular information into computational models \cite{rao2021exploring,jiang2021computational}. Motivated by these developments, we argue that a comprehensive cellular representation should jointly model transcriptomic signals, protein abundance, protein structural properties, and subcellular localization.

In this work, we propose a cross-attention-based multimodal framework for subcellular cell representation learning. Our model integrates RNA expression matrices with protein features and explicitly incorporates protein structural information to capture functional molecular constraints. Furthermore, we introduce subcellular localization as an additional modality to encode intracellular spatial organization. By jointly modeling molecular abundance, structural biology, and spatial context, our framework learns biologically informed cellular embeddings that better reflect the complexity of cellular systems and enable more accurate downstream biological inference.
\section*{Material \& Methods}


\subsection*{Preprocessing and Pre-training}

To construct a unified multimodal representation space for RNA and protein data, we first perform systematic preprocessing on both modalities to ensure consistency, quality, and biological interpretability.

The scRNA-seq data is preprocessed using the strategy widely adopted in previous studies \cite{setty2019characterization, bergen2020generalizing}. After filtering out genes with fewer than 20 expressed counts, the top 3,000 highly variable genes are selected. 
The gene expression counts are then normalized and log transformed. In detail, each cell's total counts are scaled to a fixed target sum (e.g., 10,000), as in \( x_{ij}^{norm}=\frac{x_{ij}}{\sum_{j}x_{ij}}*10000 \). Here, \( x_{ij} \) denotes the expression value of gene \( j \) in cell \( i \). The log transformation, applied as \( x_{ij}^{log}= \log(1 + x_{ij}^{norm}) \), is widely used in the preprocessing of scRNA-seq data analysis methods to reduce the dynamic range and attenuate extreme values and preserve relative differences among lowly expressed genes \cite{street2018slingshot, trapnell2017monocle, setty2019characterization, li2024relay, li2024tfvelo}.

For protein data, we collect protein sequences and associated annotations from UniProt \cite{uniprot2015uniprot}. Each protein is represented not only by its amino acid sequence but also by structural and functional annotations when available. To capture structural information, we optionally incorporate embeddings derived from pretrained protein structure models or sequence-based encoders. This allows the model to encode both sequence-level and structure-aware representations of proteins.

To establish a consistent mapping between RNA and protein modalities, we leverage UniProt as a biological bridge. Specifically, gene symbols from the RNA modality are first mapped to their corresponding protein entries in UniProt. This mapping enables alignment between transcript-level measurements and protein-level entities, ensuring that each gene expression feature can be associated with its corresponding protein representation when available. Through this procedure, we construct a unified RNA–protein correspondence dictionary, which serves as the foundation for subsequent multimodal fusion.

After preprocessing and alignment, both RNA and protein features are projected into a shared embedding space, enabling direct interaction between modalities in downstream cross-attention modules.

\subsection*{Protein Structure and RNA Foundation Embedding}

To obtain biologically meaningful initial representations for both protein and RNA modalities, we leverage pretrained domain-specific models that encode structural and contextual information beyond raw sequence or expression signals.

For the protein modality, we first retrieve protein sequences and annotations from UniProt \cite{uniprot2015uniprot}. To incorporate structural information, each protein is further processed using FoldExplorer\cite{shen2024foldexplorer}, a structure-aware protein representation model designed for fast and accurate protein structure search via sequence-enhanced graph embedding . FoldExplorer integrates sequence and structural priors to generate embeddings that capture both evolutionary and conformational properties of proteins. Through this process, we obtain an initial protein embedding that encodes sequence identity, structural topology, and functional similarity in a unified representation space.

Beyond sequence and structure, protein function is also strongly influenced by its subcellular localization. To explicitly incorporate spatial cellular context, we further augment the FoldExplorer-derived protein embeddings with subcellular localization priors derived from UniProt annotations. Specifically, we construct a seven-dimensional localization vector, where each dimension corresponds to the dominant probability of the protein being localized in one of the major cellular compartments, including cytoplasm, nucleus, plasma membrane, mitochondria, endoplasmic reticulum, Golgi apparatus, and endosome.

This subcellular prior is concatenated with the structural protein embedding, enabling the model to jointly capture molecular-level properties and spatial organizational context. By integrating sequence, structural, and localization information, the resulting protein representation provides a more comprehensive and biologically grounded embedding for downstream multimodal fusion.

For the RNA modality, we utilize scGPT, a generative pre-trained transformer model for single-cell transcriptomic data, to obtain contextualized gene and cell embeddings \cite{chen2023scgpt}. Specifically, gene expression matrices are first tokenized at the gene level, where each gene is associated with a unique gene identifier. These gene IDs, together with their corresponding expression values, are input into scGPT to generate latent representations that capture gene-gene dependencies and cell-state-specific transcriptional programs. This allows the RNA embedding to move beyond simple count-based representations and instead reflect higher-order regulatory relationships learned from large-scale single-cell datasets.

By aligning FoldExplorer-derived protein embeddings with scGPT-derived RNA embeddings, we construct modality-specific but biologically informed representations that serve as the foundation for subsequent cross-attention-based multimodal fusion.

\subsection*{The Framework of RNA\&Protein SubCell Embedding}
We leverage the cross-attention model\cite{lin2022cat} as the backbone of our framework. The transcriptomic module takes two types of inputs, including a prior gene knowledge, and their corresponding expression values. In parallel, the proteomic module processes protein tokens as input.

These heterogeneous inputs are first projected into modality-specific embedding spaces through dedicated embedding layers and subsequently encoded using separate transformer-based architectures. Within the transcriptomic pathway, we adopt a masked-attention transformer architecture inspired by scGPT, which is designed to capture complex gene–gene and gene–cell dependencies from partially observed expression profiles.

The resulting RNA embeddings are then used to guide a cross-attention mechanism within the proteomic encoder, enabling RNA-informed modeling of protein abundance from corresponding cell surface protein tokens. Through this cross-modal interaction, transcriptomic signals provide contextual guidance for proteomic representation learning, facilitating biologically consistent alignment between RNA and protein modalities.

Finally, this RNA-guided proteomic reasoning yields a unified multi-omic joint embedding, which integrates complementary information from both modalities and supports a wide range of downstream analytical tasks.

\subsection*{RNA\&Protein Encoder Module}

To address the non-sequential nature of single-cell transcriptomic data and to effectively leverage pretrained representations from scGPT, we adopt a masked-attention Transformer architecture to encode RNA embeddings. Given an input RNA embedding matrix
\begin{equation}
h^{(i)} \in \mathbb{R}^{M \times D},
\end{equation}
where $M$ denotes the number of genes and $D$ denotes the embedding dimension, we apply a multi-layer Transformer encoder to model gene–gene interactions.

The standard self-attention mechanism is defined as:
\begin{equation}
\text{Attention}(Q, K, V) = \text{Softmax}\left(\frac{QK^\top}{\sqrt{d}} + A_{\text{mask}}\right)V,
\end{equation}
where $Q, K, V \in \mathbb{R}^{M \times d}$ are query, key, and value matrices, and $A_{\text{mask}}$ is a task-specific attention mask.

The projection is computed as:
\begin{equation}
Q = hW_Q, \quad K = hW_K, \quad V = hW_V,
\end{equation}
where $W_Q, W_K, W_V \in \mathbb{R}^{D \times d}$ are learnable parameters.

To adapt to gene expression modeling, we introduce a masked attention\cite{cheng2022masked} strategy that enables autoregressive-style prediction on non-sequential data. Specifically, we partition genes into observed ("known") and masked ("unknown") subsets, and define the attention mask as:
\begin{equation}
A_{\text{mask}}(i,j) =
\begin{cases}
0, & \text{if gene } j \text{ is observable for } i \\
-\infty, & \text{otherwise}
\end{cases}
\end{equation}

This prevents each query token from attending to future or unknown genes, enforcing a controlled generation process.

The RNA representation after the final Transformer layer is:
\begin{equation}
h_n^{(i)} = \text{Transformer}(h^{(i)}) \in \mathbb{R}^{M \times D}.
\end{equation}

To improve computational efficiency for large-scale gene panels (where $M$ can reach tens of thousands), we adopt FlashAttention\cite{dao2022flashattention}, a memory-efficient implementation of self-attention that reduces quadratic memory complexity while maintaining exact attention computation.

For the protein modality, we employ a standard Transformer encoder initialized randomly to model intra-protein relationships. Given a protein embedding matrix:
\begin{equation}
p^{(i)} \in \mathbb{R}^{P \times D},
\end{equation}
where $P$ is the number of protein tokens, we first apply stacked self-attention layers to capture structural and contextual dependencies among protein tokens.

For the $n$-th layer, the self-attention operation is defined as:
\begin{equation}
\text{SelfAttn}_n(p) = \text{Softmax}\left(\frac{Q_p K_p^\top}{\sqrt{d}}\right)V_p,
\end{equation}
where:
\begin{equation}
Q_p = pW_Q^p,\quad K_p = pW_K^p,\quad V_p = pW_V^p.
\end{equation}

Each Transformer block is followed by residual connections, layer normalization, and a position-wise feed-forward network, producing the final intra-protein representation:
\begin{equation}
\tilde{p}^{(i)} \in \mathbb{R}^{P \times D}.
\end{equation}

To integrate transcriptomic context into protein representation learning, we introduce a cross-attention mechanism that fuses RNA and protein modalities.

Specifically, RNA embeddings are used as queries and keys, while protein embeddings serve as values:
\begin{equation}
Q_r = h_n W_Q^c,\quad K_r = h_n W_K^c,\quad V_p = \tilde{p} W_V^c.
\end{equation}

The cross-attention operation is then defined as:
\begin{equation}
\text{CrossAttn}(h_n, \tilde{p}) =
\text{Softmax}\left(
\frac{Q_r K_p^\top}{\sqrt{d}}
\right)V_p.
\label{eq:cross_attention}
\end{equation}

This produces an RNA-conditioned protein representation:
\begin{equation}
z^{(i)} \in \mathbb{R}^{P \times d}.
\end{equation}

The output is then projected back to the original embedding dimension:
\begin{equation}
\hat{z}^{(i)} = z^{(i)} W_O + \tilde{p}^{(i)},
\end{equation}
followed by residual connection and layer normalization.

The resulting embedding:
\begin{equation}
\hat{z}^{(i)} \in \mathbb{R}^{P \times D}
\end{equation}
encodes both transcriptomic context and proteomic structural information, and is used as the final multimodal representation for downstream tasks.

\subsection*{Cell Representation Learning Via Cross-Modal Fusion}

To construct a unified multi-modal cell representation, we design a hierarchical embedding framework that integrates protein-level structure modeling with RNA-guided cross-modal fusion.

For the protein modality, we prepend a special classification token $\langle \mathrm{cls} \rangle$ to the sequence of protein tokens:
\begin{equation}
\langle \mathrm{cls} \rangle, P_1, P_2, \dots, P_N,
\end{equation}
where $N$ denotes the number of protein tokens. This design allows the model to aggregate global protein-level information into a single latent representation.

After passing through multiple layers of self-attention, the protein sequence is transformed into:
\begin{equation}
\langle \mathrm{cls} \rangle', P_1', P_2', \dots, P_N',
\end{equation}
where $\langle \mathrm{cls} \rangle'$ serves as a global protein representation that integrates intra-protein structural and contextual dependencies learned via self-attention.

To further integrate transcriptomic information, we introduce a cross-attention-based fusion module. In this module, protein representations are used as queries, while RNA embeddings serve as keys and values, enabling RNA-guided modulation of protein representations:
\begin{equation}
Q_p = \tilde{P}W_Q,\quad K_r = HW_K,\quad V_r = HW_V,
\end{equation}
where $\tilde{P}$ denotes the protein representation after self-attention, and $H$ denotes the RNA embedding.

The cross-attention operation is defined as Eq.~\eqref{eq:cross_attention}. After fusion, we obtain a multi-modal representation:
\begin{equation}
\tilde{P}_{\text{fusion}} \in \mathbb{R}^{(N+1) \times D},
\end{equation}
where each protein token is enriched with transcriptomic context while preserving structural information from the protein encoder.

Finally, we extract the hidden state corresponding to the $\langle \mathrm{cls} \rangle$ token from the fusion block:
\begin{equation}
z_{\text{cell}} = \tilde{P}_{\text{fusion}}[\langle \mathrm{cls} \rangle],
\end{equation}
which serves as the final multi-modal cell representation. This embedding simultaneously encodes:

\begin{itemize}
\item intra-protein structural organization (from self-attention),
\item transcriptomic context (from cross-attention),
\item and global cellular state (via $\langle \mathrm{cls} \rangle$ pooling).
\end{itemize}

\subsection*{Loss Function}

The proposed model is pretrained using a composite objective function designed to learn biologically meaningful multimodal representations from single-cell data. The overall training objective integrates two complementary components: (1) a masked gene expression modeling task within the RNA modality; (2) cross-modal prediction of protein expression conditioned on transcriptomic features. Together, these objectives enable the model to capture both intra-modal gene regulatory structures and inter-modal transcript–protein relationships in a unified framework.

\subsubsection*{Masked Gene Expression Modeling}

To learn gene–gene dependencies from partially observed transcriptomic profiles, we adopt a masked gene expression (MGE) modeling objective, inspired by masked language modeling in Transformer-based architectures.

Given a cell $i$ with gene expression vector $x^{(i)} \in \mathbb{R}^{M}$, we randomly mask a subset of genes $\mathcal{U}^{(i)}_{\text{unk}} \subset \{1, \dots, M\}$ during training. 

For each masked gene $j$, a prediction head (MLP) maps its hidden representation $h^{(i)}_j$ to the reconstructed expression value:
\begin{equation}
\hat{x}^{(i)}_j = \mathrm{MLP}(h^{(i)}_j).
\end{equation}

The MGE objective minimizes the mean squared error between predicted and observed expression values:
\begin{equation}
\mathcal{L}_{\text{MGE}} =
\frac{1}{|\mathcal{U}^{(i)}_{\text{unk}}|}
\sum_{j \in \mathcal{U}^{(i)}_{\text{unk}}}
\left( x^{(i)}_j - \hat{x}^{(i)}_j \right)^2.
\end{equation}

Optimizing this objective encourages the model to learn gene–gene co-expression structure and latent regulatory dependencies from incomplete transcriptomic observations.

\subsubsection*{Cross-Modal Protein Prediction Loss}

To align transcriptomic and proteomic modalities, we introduce a cross-modal prediction objective that estimates protein expression from RNA-derived representations.

Let $p^{(i)} \in \mathbb{R}^{P}$ denote the protein expression vector for cell $i$. Given the RNA embedding $h^{(i)}$, the model predicts protein abundance via a cross-attention-based fusion module followed by a regression head:
\begin{equation}
\hat{p}^{(i)} = f_{\theta}(h^{(i)}, p^{(i)}),
\end{equation}
where $f_{\theta}(\cdot)$ denotes the multimodal fusion network.

The corresponding loss is defined as:
\begin{equation}
\mathcal{L}_{\text{cross}} =
\| p^{(i)} - \hat{p}^{(i)} \|_2^2.
\end{equation}

This objective enforces consistency between transcriptomic and proteomic representations and facilitates biologically grounded cross-modal alignment.

\subsubsection*{Overall Objective}

The final training objective is a weighted combination of the three losses:
\begin{equation}
\mathcal{L} =
\lambda_1 \mathcal{L}_{\text{MGE}} +
\lambda_2 \mathcal{L}_{\text{cross}}
\end{equation}
where $\lambda_1 = 0.6$ and $\lambda_2 = 0.4$ control the contributions of the masked gene expression modeling loss and the cross-modal prediction loss, respectively.

\newpage
\section*{Results}

\subsection*{Multimodal RNA--Protein representations improve cell type identification in PBMC datasets}

To evaluate the effectiveness of our multimodal cell embedding framework for cell type identification, 
we applied our model to PBMC datasets\cite{hao2021integrated} containing diverse immune cell populations. 
The learned embeddings were visualized using UMAP and compared with experimentally annotated cell identities 
(Fig.2a). The multimodal representations generated by our model successfully preserved the 
global organization of immune populations, including B cells, T cells, monocytes, natural killer (NK) cells, 
dendritic cells, and platelets. The predicted cell type distribution showed strong concordance with the 
original annotations, demonstrating that our model effectively captures biologically meaningful cellular 
structures.

The UMAP visualization further demonstrated that our model was capable of separating closely related immune 
subpopulations. In particular, distinct T cell states, including CD4 naive, CD4 TCM, CD4 TEM, CD8 naive, 
CD8 TCM, and CD8 TEM cells, were well resolved in the embedding space. These results indicate that 
integrating transcriptomic and protein-level information enables the model to capture subtle phenotypic 
differences that are difficult to distinguish using transcriptomic information alone.

To quantitatively evaluate classification performance, we calculated the confusion matrix across all immune 
cell subtypes (Fig.2b). The majority of cell populations showed strong diagonal enrichment, 
indicating high consistency between predicted and annotated labels. Major immune populations, including 
CD14 monocytes, CD16 monocytes, naive and memory B cells, NK cells, and T cell populations, achieved high 
classification accuracy with minimal misclassification. Moreover, the model demonstrated robust performance 
in distinguishing biologically similar cell states, highlighting the advantage of multimodal representation 
learning for fine-grained immune cell annotation.

We further benchmarked our approach against representative single-cell representation and integration methods, including the transcriptomic foundation model \textit{scGPT} and the widely used conventional integration framework Seurat\cite{gribov2010seurat}, using four commonly used evaluation metrics: accuracy, precision, recall, and macro-F1 score (Fig.~2c). Cell-type labels were obtained from curated annotations provided by the original datasets and were used as reference labels for evaluation.

On the 10x Multiome PBMC dataset, our method achieved an accuracy of $0.962$, precision of $0.961$, recall of $0.937$, and macro-F1 score of $0.947$. Compared with \textit{scGPT}, our approach achieved consistent improvements across most evaluation metrics. These improvements indicate that incorporating complementary protein-derived features and subcellular-scale information into transcriptomic representations provides additional biological signals for distinguishing closely related cell populations.

Furthermore, on the multi-tissue immune cell benchmark, our model achieved accuracy, precision, recall, and 
macro-F1 scores of $0.922$, $0.898$, $0.920$, and $0.905$, respectively. These results substantially 
outperformed Seurat, which obtained scores of $0.790$, $0.848$, $0.732$, and $0.761$, respectively. 
The improved recall and macro-F1 scores indicate that our multimodal representation provides better 
generalization capability across heterogeneous immune cell populations.

Overall, these results demonstrate that explicitly modeling RNA--protein interactions through cross-modal 
attention enables the construction of more informative cell embeddings, leading to accurate cell type 
identification and robust transferability across diverse single-cell datasets.



\subsection*{Effect of Subcellular Cell Embedding on Cell Type Identification}

To investigate the contribution of subcellular-scale information to cell representation learning, we performed an ablation analysis by comparing models with and without the incorporation of subcellular embeddings. Although the introduced subcellular representation consists of only seven dimensions, it provides complementary biological information beyond conventional transcriptomic features. As shown in the confusion matrix analysis (Fig.3b), incorporating subcellular embeddings improved the discrimination of several biologically related cell populations. In particular, the classification accuracy of CD4 TCM cells was enhanced, accompanied by a reduction in misclassification between CD4 TCM and cDC2 populations, suggesting that subcellular localization information contributes to resolving subtle cellular heterogeneity(Fig.4b). The improvement in CD4$^{+}$ TCM identification may be attributed to the incorporation of subcellular-aware features. CD4$^{+}$ TCM cells possess distinct intracellular states characterized by enhanced mitochondrial fitness and specialized metabolic programs required for long-term persistence and rapid immune recall responses.\cite{rose2023distinct} Moreover, T cell activation and differentiation are accompanied by dynamic redistribution of signaling proteins among membrane, cytoplasmic, and nuclear compartments.\cite{joshi2019tcellsubc} Therefore, integrating subcellular localization information provides complementary biological cues beyond transcript abundance, facilitating the resolution of subtle heterogeneity among immune cell populations.

Furthermore, the model incorporating subcellular-scale embeddings consistently achieved improved performance across multiple evaluation metrics compared with the model without subcellular information (Fig.3c). On the 10x Multiome PBMC dataset, the incorporation of subcellular embeddings improved the overall classification performance, increasing the accuracy from 0.92 to 0.94 and the macro-F1 score from 0.90 to 0.92. Similar improvements were observed on the multi-tissue immune cell dataset, demonstrating that subcellular representations provide robust and transferable biological signals across diverse cellular contexts.

These results indicate that even a compact seven-dimensional subcellular embedding can effectively complement transcriptomic information, enabling more informative cell representations and enhancing downstream cell type annotation.

\subsection*{Cross-modal Representation Learning Enables Robust Batch Integration while Preserving Cellular Identity}

Batch effects arising from donor variability, experimental conditions, and sequencing platforms represent a major challenge in single-cell analysis. Effective integration requires removing technical variations while preserving biologically meaningful differences between cell populations. However, existing integration approaches primarily rely on transcriptomic information, which may be insufficient for distinguishing biological signals from batch-associated fluctuations, particularly in heterogeneous and multi-modal datasets.

To evaluate the ability of our model to learn batch-robust cellular representations, we performed integration experiments across multiple heterogeneous single-cell datasets containing diverse biological backgrounds and technical conditions. Our framework jointly models RNA expression, protein abundance, and subcellular-scale information through a cross-modal attention mechanism, enabling the model to capture shared biological characteristics across datasets while reducing modality-specific and batch-specific variations.

We first evaluated batch integration performance on multi-batch scRNA-seq datasets collected from different donors. As summarized in Table~\ref{tab:integration_benchmark}, our method achieved consistently improved biological conservation compared with existing approaches. On the Human PBMCs Covid dataset, our model obtained the highest AvgBIO score (0.657), together with the highest NMI$_{cell}$ (0.742) and ARI$_{cell}$ (0.691), outperforming scGPT (AvgBIO = 0.626), TotalVI (0.607), and Scanpy (0.519). These results indicate that our learned representations better preserve cell-type-specific biological structures while effectively integrating cells across heterogeneous batches.

We further evaluated the integration capability on a more complex multi-modal benchmark consisting of TEA-seq, ECCITE-seq, and CITE-seq datasets. By jointly incorporating transcriptomic, proteomic, and subcellular information, our method achieved superior overall integration performance, with an AvgBIO score of 0.567 and an Overall score of 0.728. Compared with scGPT, our approach improved biological conservation (AvgBIO: 0.567 vs. 0.517), cell-type clustering consistency (NMI$_{cell}$: 0.638 vs. 0.548; ARI$_{cell}$: 0.408 vs. 0.339), and batch correction performance (AvgBATCH: 0.968 vs. 0.951). In particular, the improvements in Graph Connectivity (0.992 vs. 0.988) and ASW$_{batch}$ (0.943 vs. 0.913) demonstrate that our model effectively aligns cells from different experimental conditions while maintaining biologically meaningful structures.

Visualization of the learned embeddings further confirmed these quantitative improvements (Fig.4). Compared with conventional integration approaches, our method generated a more coherent latent space in which cells from different batches were effectively aligned while retaining distinct immune cell identities. The UMAP visualization showed clear separation of major immune populations, including B cells, T cell subsets, monocyte populations, and NK cells, demonstrating that the learned representation preserves both global lineage structure and fine-grained cellular heterogeneity.

To further quantify biological preservation, we calculated the biological conservation score (AvgBIO) across different integration methods (Fig.4A). Our method achieved the highest AvgBIO score (0.795), exceeding scGPT (0.723), Harmony (0.627), and Seurat (0.541). This improvement suggests that incorporating additional biological modalities provides complementary information beyond transcriptomic representations alone, enabling more accurate preservation of cellular identity during integration.

We further investigated the contribution of multimodal information to representation learning. The incorporation of protein features provided complementary phenotypic information beyond RNA expression, improving the resolution of closely related immune populations that exhibit overlapping transcriptional profiles. Moreover, the integration of subcellular-scale embeddings further enhanced representation robustness by introducing biologically informative constraints associated with cellular organization and molecular localization patterns. Unlike transcriptomic signals that may vary substantially across experimental batches, subcellular information provides an additional biological context that helps stabilize cell-state representations.

In particular, the improved representation quality enhanced the discrimination of closely related immune populations. For example, incorporation of subcellular localization information improved the identification of CD4$^{+}$ TCM cells and reduced confusion with transcriptionally similar populations such as cDC2, suggesting that subcellular-aware representations capture additional intracellular organization and functional state information beyond gene expression patterns.

Together, these results demonstrate that cross-modal representation learning integrating RNA, protein, and subcellular-scale information enables robust batch correction while preserving biologically meaningful cellular variation across heterogeneous single-cell datasets.
\begin{table*}[ht]
\centering
\caption{
Benchmarking single-cell integration performance across heterogeneous datasets. 
Methods are evaluated from two complementary aspects: biological conservation and batch correction.
Biological conservation is assessed using AvgBIO, which integrates normalized mutual information (NMI$_{cell}$), adjusted Rand index (ARI$_{cell}$), and average silhouette width (ASW$_{cell}$) based on cell-type annotations.
Batch correction performance is evaluated using AvgBATCH, combining ASW$_{batch}$ and graph connectivity metrics.
Higher values indicate better performance.
The best-performing method in each dataset is highlighted in bold.
}
\label{tab:integration_benchmark}

\resizebox{\textwidth}{!}{
\begin{tabular}{llcccccccc}
\toprule
\multirow{2}{*}{Dataset} &
\multirow{2}{*}{Model} &
\multicolumn{4}{c}{Biological Conservation} &
\multicolumn{4}{c}{Batch Correction} \\

\cmidrule(lr){3-6}
\cmidrule(lr){7-10}

&
&
AvgBIO &
NMI$_{cell}$ &
ARI$_{cell}$ &
ASW$_{cell}$ &
AvgBATCH &
ASW$_{batch}$ &
GraphConn &
Overall
\\
\midrule

\multirow{4}{*}{Human PBMCs Covid}

& Our method (fine-tuned)
& \textbf{0.657}
& \textbf{0.742}
& \textbf{0.691}
& 0.538
& -
& -
& -
& -
\\

& scGPT (fine-tuned)
& 0.626
& 0.710
& 0.608
& \textbf{0.545}
& -
& -
& -
& -
\\

& TotalVI
& 0.607
& 0.689
& 0.608
& 0.522
& -
& -
& -
& -
\\

& Scanpy
& 0.519
& 0.589
& 0.475
& 0.494
& -
& -
& -
& -
\\

\midrule

\multirow{2}{*}{TEA-seq+ECCITE-seq+CITE-seq}

& Our method (fine-tuned)
& \textbf{0.567}
& \textbf{0.638}
& \textbf{0.408}
& 0.657
& \textbf{0.968}
& \textbf{0.943}
& \textbf{0.992}
& \textbf{0.728}
\\

& scGPT (fine-tuned)
& 0.517
& 0.548
& 0.339
& \textbf{0.664}
& 0.951
& 0.913
& 0.988
& 0.690
\\

\bottomrule

\end{tabular}
}
\end{table*}
\FloatBarrier
\section*{Discussion}

In this work, we present a multimodal framework for learning biologically informed single-cell representations by jointly integrating transcriptomic and proteomic information. Unlike existing single-cell foundation models that primarily rely on gene expression profiles, our approach explicitly incorporates protein-level structure and subcellular spatial context, enabling a more comprehensive characterization of cellular states.

Specifically, we propose a cross-attention-based architecture that aligns RNA and protein modalities within a unified representation space. On the protein side, we leverage structure-aware embeddings derived from FoldExplorer, enriched with subcellular localization priors from UniProt annotations. On the RNA side, we adopt a pretrained Transformer model inspired by scGPT to capture gene regulatory dependencies from high-dimensional expression data. By bridging these two modalities through cross-attention mechanisms, our model learns biologically grounded interactions between transcriptomic programs and proteomic functions.

Furthermore, we design a CLS-token-based fusion strategy to obtain a compact yet expressive cell-level representation, which simultaneously encodes intra-protein structural information, transcriptomic context, and spatial cellular organization. The proposed framework is trained using a composite objective that combines masked gene modeling and cross-modal prediction, enabling both intra-modal structure learning and inter-modal alignment.

Despite these advances, several limitations remain. First, the current model relies on existing protein annotations and structure-derived embeddings, which may not fully capture dynamic conformational changes or context-specific protein behavior. Second, subcellular localization is represented as a coarse-grained prior, and more fine-grained spatial modeling could further improve biological fidelity. Third, the current framework focuses on static snapshots of cells and does not explicitly model temporal dynamics such as differentiation trajectories or perturbation responses.

In future work, we envision extending this framework in several directions. First, integrating dynamic protein structure representations and spatially resolved omics data could provide a more detailed view of intracellular organization. Second, incorporating perturbation-aware modeling would enable the simulation and prediction of cellular responses under genetic or pharmacological interventions, paving the way toward virtual cell modeling. Finally, scaling the framework to larger multi-modal datasets and more diverse biological conditions may further enhance its generalization ability and applicability in biomedical research.



\newpage


\section*{RESOURCE AVAILABILITY}


\subsection*{Lead contact}


Requests for further information and resources should be directed to and will be fulfilled by the lead contact, Hongbin Shen (hbshen@sjtu.edu.cn).

\subsection*{Data and code availability}


The code to reproduce results, together with documentation and examples of usage, is available on GitHub at \href{https://github.com/ZHOUZHEN2002/RNA_Protein/tree/master}{https://github.com/ZHOUZHEN2002/RNA\_Protein/tree/master}

\section*{ACKNOWLEDGMENTS}


Acknowledgements
This work was supported by the National Natural Science Foundation of China (No. 62573293, 62473257), and the Science and Technology Commission of Shanghai Municipality (No. 24ZR1435300, 24510714300).

\section*{AUTHOR CONTRIBUTIONS}


Author contributions:  H.S. and Z.Z. designed research; Z.Z. performed research; J.L. and X.P. contributed new reagents/analytic tools; Z.Z. analyzed data; and Z.Z., Y.L., J.L. and H.S. wrote the paper.

\section*{DECLARATION OF INTERESTS}


The authors declare no competing interests.

\section*{DECLARATION OF GENERATIVE AI AND AI-ASSISTED TECHNOLOGIES}


During the preparation of this manuscript, GPT-5 was used solely to assist with language editing and improve the clarity and readability of the text. All scientific content, data analysis, interpretation, and conclusions were developed and verified by the authors, who take full responsibility for the final manuscript.







\section*{MAIN FIGURE TITLES AND LEGENDS}

\noindent\includegraphics[width=1\linewidth]{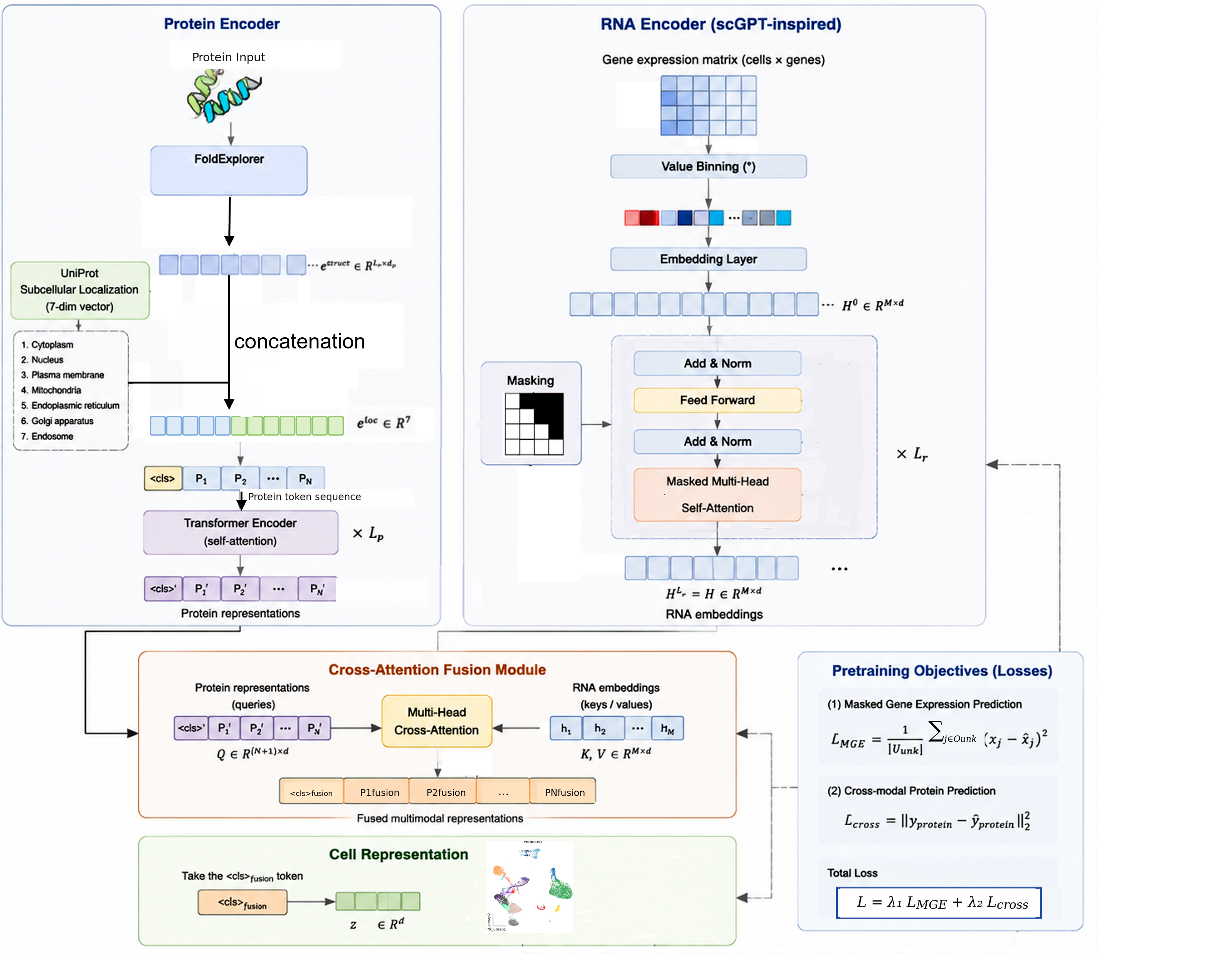}
\subsection*{Figure 1. Overview of the proposed multimodal framework for subcellular cell representation learning.}

The model integrates protein structure-aware representations and RNA transcriptomic embeddings through a unified cross-attention architecture to generate biologically informed cell representations.

Left: Protein encoding module. Protein sequences and annotations are retrieved from UniProt and encoded using a structure-aware embedding model (FoldExplorer), which integrates sequence and structural priors to produce protein representations. Subcellular localization information is further incorporated as a 7-dimensional vector capturing major cellular compartments, including cytoplasm, nucleus, plasma membrane, mitochondria, endoplasmic reticulum, Golgi apparatus, and endosome. These features are fused and processed through a Transformer encoder with self-attention layers to obtain protein-level representations.

Right: RNA encoding module. Single-cell gene expression matrices are processed using a masked-attention Transformer inspired by scGPT. The model learns gene–gene dependencies via masked gene modeling, enabling robust reconstruction of missing gene expression values. FlashAttention is used to improve computational efficiency for large gene sets.

Middle: Cross-modal fusion. RNA embeddings are used to guide protein representation learning through a cross-attention mechanism, where transcriptomic signals act as contextual information for proteomic encoding. The fused representations are aggregated using a CLS token to generate the final multi-modal cell embedding, which captures transcriptomic, proteomic, structural, and spatial information in a unified latent space.

\newpage
\noindent\includegraphics[width=1\linewidth]{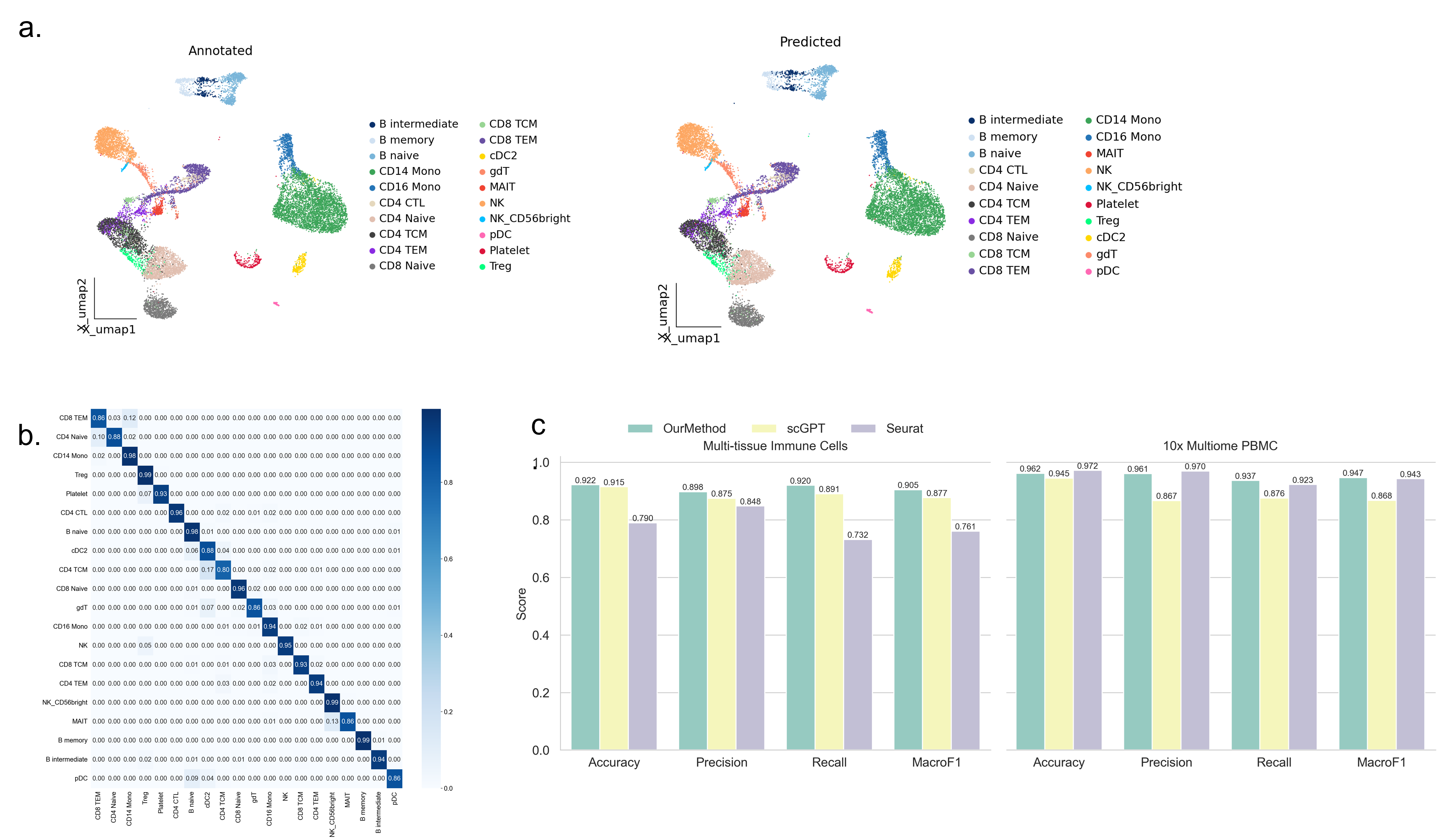}
\subsection*{Figure 2. Cross-modal RNA--protein representation learning improves PBMC cell type identification.}
(a) UMAP visualization of PBMC cells embedded by the proposed RNA--protein fusion model. 
The left panel shows manually annotated cell populations, whereas the right panel displays predicted cell identities 
based on the learned multimodal embeddings.

(b) Confusion matrix illustrating classification accuracy across immune cell subtypes, demonstrating the capability 

of the model to distinguish closely related immune populations.
(c) Comparison of cell annotation performance between our method, \textit{scGPT}, and Seurat using four evaluation metrics 
(accuracy, precision, recall, and macro-F1) on a multi-tissue immune cell benchmark and the 10x Multiome PBMC dataset.

\newpage
\noindent\includegraphics[width=1\linewidth]{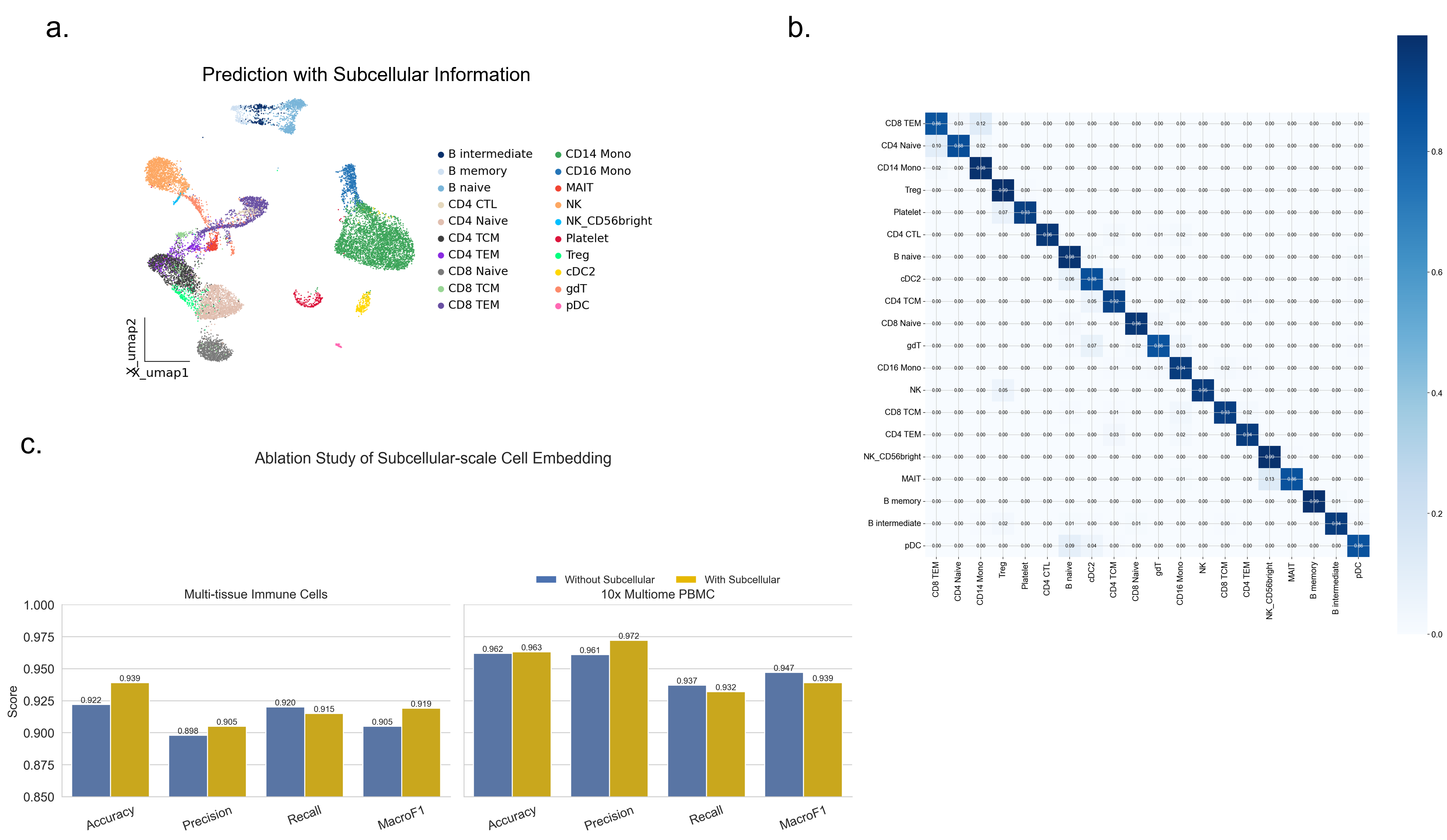}
\subsection*{Figure 3. Effect of subcellular-scale cell embedding on cell representation and cell type identification.}

(a) UMAP visualization of PBMC cells based on the learned cell embeddings incorporating subcellular-scale information. 
Cells are colored according to manually annotated cell types, demonstrating the preservation of diverse immune cell populations and the ability of the proposed representation to capture biologically meaningful cellular heterogeneity.

(b) Confusion matrix of cell type classification performance using the subcellular-enhanced cell embeddings. 
The model achieves high classification accuracy across major immune populations and effectively resolves closely related cell types, indicating that subcellular information provides complementary signals for improving cell identity discrimination.

(c) Ablation analysis evaluating the contribution of subcellular-scale embeddings. 
The performance of models with and without the incorporation of subcellular embeddings is compared on the 10x Multiome PBMC dataset and the multi-tissue immune cell benchmark using four evaluation metrics, including accuracy, precision, recall, and macro-F1 score. 
Despite consisting of only seven dimensions, the compact subcellular representation consistently improves cell type annotation performance, demonstrating that subcellular-scale information provides additional biological context beyond transcriptomic features and enhances cell representation learning.
\clearpage
  \begin{figure}[p]
      \centering
      \includegraphics[
          width=\linewidth,
          height=0.9\textheight,
          keepaspectratio
      ]{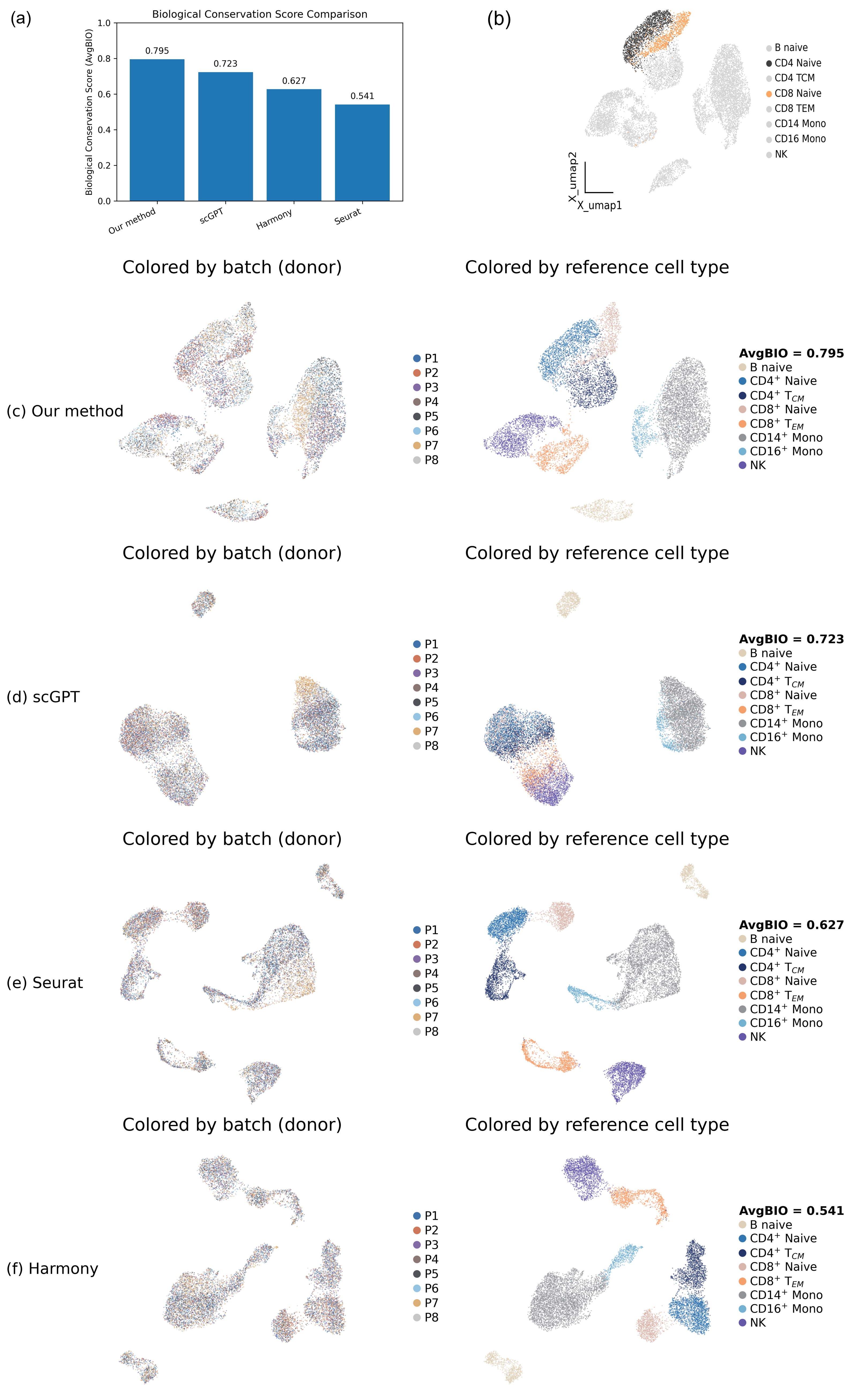}
      \label{fig:pbmc-integration}
  \end{figure}
\clearpage
\subsection*{Figure 4. UMAP visualization of the healthy PBMC dataset after integration using different methods.}
(a) Quantitative comparison of biological conservation scores (AvgBIO) among different methods. 
Our method achieves the highest biological conservation score (AvgBIO = 0.795), outperforming 
\textit{scGPT} (0.723), \textit{Harmony} (0.627), and \textit{Seurat} (0.541), indicating improved 
preservation of intrinsic biological structures during representation learning. 

(b) UMAP visualization of cell embeddings generated by our method. 
Cells are colored according to curated immune cell identities, including B naive, 
CD4$^{+}$ Naive, CD4$^{+}$ TCM, CD8$^{+}$ Naive, CD8$^{+}$ TEM, CD14$^{+}$ Mono, 
CD16$^{+}$ Mono, and NK cells. 
The learned representation enables clear separation of major immune populations while 
preserving biologically meaningful relationships among related cell states. 
In particular, incorporating subcellular localization information improves the discrimination 
of CD4$^{+}$ TCM cells and reduces their confusion with closely related cDC2 populations, suggesting that subcellular-aware representations facilitate the resolution of subtle cellular heterogeneity. 

(c–f) UMAP visualizations of the healthy PBMC dataset after integration using (c) Our method, (d) scGPT, (e) Seurat, and (f) Harmony. In each panel, the left UMAP
is colored by batch (donor; P1–P8) to assess batch mixing, whereas the right UMAP shows the same cells colored by reference cell-type annotations to evaluate the
preservation of biological identity. Effective integration is reflected by the mixing of donor labels within shared embedding regions while maintaining coherent
separation of known immune cell populations. Compared with conventional integration methods, our approach provides improved alignment between the learned
embedding structure and biological cell identities, achieving the highest AvgBIO score (0.795), compared with scGPT (0.723), Seurat (0.627), and Harmony (0.541).
AvgBIO values are displayed above the cell-type legends.

\newpage


\bibliography{references}

@article{theodoris2023geneformer,
  title={Transfer learning enables predictions in network biology},
  author={Theodoris, Christina V and Xiao, Ling and Chopra, Ashish and others},
  journal={Nature},
  year={2023}
}

@article{chen2023scgpt,
  title={scGPT: toward building a foundation model for single-cell multi-omics using generative AI},
  author={Chen, Zhiyuan and others},
  journal={bioRxiv},
  year={2023}
}

@article{jumper2021alphafold,
  title={Highly accurate protein structure prediction with AlphaFold},
  author={Jumper, John and Evans, Richard and Pritzel, Alexander and others},
  journal={Nature},
  volume={596},
  pages={583--589},
  year={2021}
}

@article{alberts2017molecular,
  title={Molecular Biology of the Cell},
  author={Alberts, Bruce and others},
  journal={Garland Science},
  year={2017}
}

@article{setty2019characterization,
  title={Characterization of cell fate probabilities in single-cell data with Palantir},
  author={Setty, Manu and Kiseliovas, Vaidotas and Levine, Jacob and Gayoso, Adam and Mazutis, Linas and Pe’Er, Dana},
  journal={Nature biotechnology},
  volume={37},
  number={4},
  pages={451--460},
  year={2019},
  publisher={Nature Publishing Group US New York},
note={URL: \url{https://www.nature.com/articles/s41587-019-0068-4} (accessed May 14, 2025)}
}

@article{bergen2020generalizing,
  title={Generalizing RNA velocity to transient cell states through dynamical modeling},
  author={Bergen, Volker and Lange, Marius and Peidli, Stefan and Wolf, F Alexander and Theis, Fabian J},
  journal={Nature biotechnology},
  volume={38},
  number={12},
  pages={1408--1414},
  year={2020},
  publisher={Nature Publishing Group US New York},
note={URL: \url{https://www.nature.com/articles/s41587-020-0591-3} (accessed May 14, 2025)}
}

@article{street2018slingshot,
  title={Slingshot: cell lineage and pseudotime inference for single-cell transcriptomics},
  author={Street, Kelly and Risso, Davide and Fletcher, Russell B and Das, Diya and Ngai, John and Yosef, Nir and Purdom, Elizabeth and Dudoit, Sandrine},
  journal={BMC genomics},
  volume={19},
  pages={1--16},
  year={2018},
  publisher={Springer},
note={URL: \url{https://link.springer.com/article/10.1186/s12864-018-4772-0} (accessed May 14, 2025)}
}

@article{trapnell2017monocle,
  title={Monocle: Cell counting, differential expression, and trajectory analysis for single-cell RNA-Seq experiments},
  author={Trapnell, Cole and Cacchiarelli, Davide and Qiu, Xiaojie},
  journal={Bioconductor. https://www. bioconductor. org/packages/release/bioc/html/monocle. html},
  year={2017},
note={URL: \url{https://bioconductor.statistik.tu-dortmund.de/packages/3.5/bioc/vignettes/monocle/inst/doc/monocle-vignette.pdf} (accessed May 14, 2025)}
}

@article{li2024tfvelo,
  title={TFvelo: gene regulation inspired RNA velocity estimation},
  author={Li, Jiachen and Pan, Xiaoyong and Yuan, Ye and Shen, Hong-Bin},
  journal={Nature Communications},
  volume={15},
  number={1},
  pages={1387},
  year={2024},
  publisher={Nature Publishing Group UK London},
note={URL: \url{https://www.nature.com/articles/s41467-024-45661-w} (accessed May 14, 2025)}
}

@article{li2024relay,
  title={A relay velocity model infers cell-dependent RNA velocity},
  author={Li, Shengyu and Zhang, Pengzhi and Chen, Weiqing and Ye, Lingqun and Brannan, Kristopher W and Le, Nhat-Tu and Abe, Jun-ichi and Cooke, John P and Wang, Guangyu},
  journal={Nature biotechnology},
  volume={42},
  number={1},
  pages={99--108},
  year={2024},
  publisher={Nature Publishing Group US New York},
note={URL: \url{https://www.nature.com/articles/s41587-023-01728-5} (accessed May 14, 2025)}
}

@article{shen2024foldexplorer,
  title={FoldExplorer: Fast and Accurate Protein Structure Search with Sequence-Enhanced Graph Embedding},
  author={Shen, Hong-Bin and Liu, Yuan and Zhang, Ying and Zhou, Zhen},
  year={2024}
}

@article{uniprot2015uniprot,
  title={UniProt: a hub for protein information},
  author={UniProt Consortium},
  journal={Nucleic acids research},
  volume={43},
  number={D1},
  pages={D204--D212},
  year={2015},
  publisher={Oxford University Press}
}

@inproceedings{lin2022cat,
  title={Cat: Cross attention in vision transformer},
  author={Lin, Hezheng and Cheng, Xing and Wu, Xiangyu and Shen, Dong},
  booktitle={2022 IEEE international conference on multimedia and expo (ICME)},
  pages={1--6},
  year={2022},
  organization={IEEE}
}

@inproceedings{cheng2022masked,
  title={Masked-attention mask transformer for universal image segmentation},
  author={Cheng, Bowen and Misra, Ishan and Schwing, Alexander G and Kirillov, Alexander and Girdhar, Rohit},
  booktitle={Proceedings of the IEEE/CVF conference on computer vision and pattern recognition},
  pages={1290--1299},
  year={2022}
}

@inproceedings{dao2022flashattention,
  title={Flashattention: Fast and memory-efficient exact attention with io-awareness},
  author={Dao, Tri and Fu, Daniel Y and Ermon, Stefano and Rudra, Atri and R{\'e}, Christopher},
  booktitle={Advances in neural information processing systems},
  year={2022}
}

@article{gribov2010seurat,
  title={SEURAT: visual analytics for the integrated analysis of microarray data},
  author={Gribov, Alexander and Sill, Martin and L{\"u}ck, Sonja and R{\"u}cker, Frank and D{\"o}hner, Konstanze and Bullinger, Lars and Benner, Axel and Unwin, Antony},
  journal={BMC medical genomics},
  volume={3},
  number={1},
  pages={21},
  year={2010},
  publisher={Springer}
}

@article{joshi2019tcellsubc,
  title={TcellSubC: an atlas of the subcellular proteome of human T cells},
  author={Joshi, Rubin Narayan and Stadler, Charlotte and Lehmann, Robert and Lehti{\"o}, Janne and Tegn{\'e}r, Jesper and Schmidt, Angelika and Vesterlund, Mattias},
  journal={Frontiers in immunology},
  volume={10},
  pages={2708},
  year={2019},
  publisher={Frontiers Media SA}
}

@article{rose2023distinct,
  title={Distinct transcriptomic and epigenomic modalities underpin human memory T cell subsets and their activation potential},
  author={Rose, James R and Akdogan-Ozdilek, Bagdeser and Rahmberg, Andrew R and Powell, Michael D and Hicks, Sakeenah L and Scharer, Christopher D and Boss, Jeremy M},
  journal={Communications Biology},
  volume={6},
  number={1},
  pages={363},
  year={2023},
  publisher={Nature Publishing Group UK London}
}

@article{cui2024scgpt,
  title={scGPT: toward building a foundation model for single-cell multi-omics using generative AI},
  author={Cui, Haotian and Wang, Chloe and Maan, Hassaan and Pang, Kuan and Luo, Fengning and Duan, Nan and Wang, Bo},
  journal={Nature methods},
  volume={21},
  number={8},
  pages={1470--1480},
  year={2024},
  publisher={Nature Publishing Group US New York}
}

@article{hao2024large,
  title={Large-scale foundation model on single-cell transcriptomics},
  author={Hao, Minsheng and Gong, Jing and Zeng, Xin and Liu, Chiming and Guo, Yucheng and Cheng, Xingyi and Wang, Taifeng and Ma, Jianzhu and Zhang, Xuegong and Song, Le},
  journal={Nature methods},
  volume={21},
  number={8},
  pages={1481--1491},
  year={2024},
  publisher={Nature Publishing Group US New York}
}

@article{krishna2024generalized,
  title={Generalized biomolecular modeling and design with RoseTTAFold All-Atom},
  author={Krishna, Rohith and Wang, Jue and Ahern, Woody and Sturmfels, Pascal and Venkatesh, Preetham and Kalvet, Indrek and Lee, Gyu Rie and Morey-Burrows, Felix S and Anishchenko, Ivan and Humphreys, Ian R and others},
  journal={Science},
  volume={384},
  number={6693},
  pages={eadl2528},
  year={2024},
  publisher={American Association for the Advancement of Science}
}

@article{kobayashi2022self,
  title={Self-supervised deep learning encodes high-resolution features of protein subcellular localization},
  author={Kobayashi, Hirofumi and Cheveralls, Keith C and Leonetti, Manuel D and Royer, Loic A},
  journal={Nature methods},
  volume={19},
  number={8},
  pages={995--1003},
  year={2022},
  publisher={Nature Publishing Group US New York}
}

@article{rao2021exploring,
  title={Exploring tissue architecture using spatial transcriptomics},
  author={Rao, Anjali and Barkley, Dalia and Fran{\c{c}}a, Gustavo S and Yanai, Itai},
  journal={Nature},
  volume={596},
  number={7871},
  pages={211--220},
  year={2021},
  publisher={Nature Publishing Group UK London}
}

@article{jiang2021computational,
  title={Computational methods for protein localization prediction},
  author={Jiang, Yuexu and Wang, Duolin and Wang, Weiwei and Xu, Dong},
  year={2021}
}

@article{hao2021integrated,
  title={Integrated analysis of multimodal single-cell data},
  author={Hao, Yuhan and Hao, Stephanie and Andersen-Nissen, Erica and Mauck, William M and Zheng, Shiwei and Butler, Andrew and Lee, Maddie J and Wilk, Aaron J and Darby, Charlotte and Zager, Michael and others},
  journal={Cell},
  volume={184},
  number={13},
  pages={3573--3587},
  year={2021},
  publisher={Elsevier}
}

\bigskip


\newpage

\end{document}